# Chalcogen Doping Effect on the Insulator-to-Metal Transition in GdPS

Gokul Acharya[1†], Rabindra Basnet[2†], Santosh Karki Chhetri[1], Dinesh Upreti[1], M. M. Sharma[1], Jian Wang[3], David Graf[4], Jin Hu[1,5*]

[1]*Department of Physics, University of Arkansas, Fayetteville, Arkansas 72701, USA*

[2]*Department of Physics, Morgan State University, Baltimore, Maryland 21251, USA*

[3]*Department of Chemistry and Biochemistry, Wichita State University, Wichita, Kansas 67260, USA*

[4]*National High Magnetic Field Lab, Tallahassee, FL, 32310 USA*

[5]*Institute for Nanoscience and Engineering, MonArk NSF Quantum Foundry, and Smart Ferroic Materials Center, University of Arkansas, Fayetteville, Arkansas 72701, USA*

**Abstract**

Topological semimetals offer a rich platform for exploring massless fermion physics and realizing exotic properties with potential technological applications. GdPS, a magnetic semiconductor derived from the nodal-line semimetal ZrSiS family, exhibits a field-induced insulator-to-metal transition driven by exchange splitting. This transition is accompanied by an unusual, isotropic, and gigantic negative magnetoresistance, attributed to negligible magnetic anisotropy resulting from the weak spin-orbit coupling of half-filled $Gd^{3+}$ 4$f$ orbitals and light S atoms. In this work, we investigate Se substitution, which is expected to enhance spin-orbit coupling. Indeed, we observe slightly increased magnetic anisotropy in magnetotransport. Moreover, Se substitution suppresses the field-induced insulator-to-metal transition, likely due to an enlarged band gap that demands a higher exchange splitting to close. These findings provide deeper insights into the interplay between spin-orbit coupling, magnetic anisotropy, and transport behavior in GdPS, offering guidance for future materials design for desired functionalities.

[†]equal contribution    [*]contact author: jinhu@uark.edu

**Introduction**

The discovery of topological semimetals, including Dirac and Weyl semimetals characterized by symmetry-protected linear band crossings at discrete points in momentum space, has opened exciting avenues for both fundamental research and technological innovation. These materials exhibit a range of exotic properties such as large magnetoresistance (MR), high carrier mobility, chiral anomaly, and surface Fermi arcs, making them promising platforms for exploring relativistic quasiparticles and developing advanced devices[1–4]. A distinct class of topological semimetals, the nodal line semimetals, features linear band crossings along one-dimensional loops or lines in momentum space. Among them, the ZrSiS-type compounds such as Zr(Si/Ge)(S/Se/Te)[5–11], Hf(Si/Ge)(S/Se/Te)[12–16] and ZrSnTe[17,18] have attracted significant attention. These compounds crystallize in the PbFCl-type structure featuring square nets of Si/Ge/Sn atoms, and host both three-dimensional (3D) nodal-line states near the Fermi level ($E_F$) and two-dimensional (2D) nodal-point Dirac states protected by non-symmorphic symmetry[4,10]. This rich band topology has been linked to phenomena such as enhanced electronic correlations[19], pressure-induced topological phase transitions[20,21], and unconventional surface state arising from surface symmetry reduction[22,23].

The chemical flexibility of this material system extends to substitution of Zr/Hf by magnetic lanthanide ($Ln$) elements, giving rise to $Ln$SbTe compounds with 4$f$ magnetism and Sb-based planar layers[24–52]. These $Ln$SbTe materials allow for composition non-stoichiometry with partial Te substitution for the Sb layer, leading to $Ln$Sb$_{2-x}$Te$_x$ compositions. A broad range of intriguing phenomena have been observed or predicted, such as tunable topological phases[26] with rich magnetic properties[24,26–29,41], magnetic devil's staircase[27,36], charge density wave (CDW)[36,53], and possible electronic correlation enhancement[28,29].

Beyond the *Ln*SbTe system, recent interest has focused on another derivative *Ln*PS, which features distorted P planes[54]. Unlike the square-net structures of ZrSiS and stoichiometric *Ln*SbTe, where Dirac nodal lines are protected by the $C_{2v}$ symmetry, the armchair-like distortion of the P layers in *Ln*PS breaks mirror symmetry. This distortion significantly alters the electronic structure, opening a sizable band gap of the Dirac crossings near the Fermi energy ($E_F$) and creating a semiconductor/insulator state[55]. A representative example is GdPS, an antiferromagnetic (AFM) semiconductor with a band gap of ~0.5 eV owing to symmetry-breaking[54]. Surface K deposition is found to resort the square-plane symmetry for P-layer and recover the Dirac crossing[56]. Interestingly, when subjected to a magnetic field above 15 T, GdPS undergoes a transition to a field-polarized ferromagnetic (FM) state[54]. The alignment of $Gd^{3+}$ spins under increasing field drives strong *f-d* exchange coupling, leading to a substantial exchange splitting (>0.5 eV) of the Gd *d*-band. This band shift brings the spin-polarized band across $E_F$, producing a remarkable insulator-to-metal transition accompanied by gigantic negative MR[54].

The strong exchange splitting, which results in insulator-to-metal transition and colossal negative MR, is caused by strong magnetization from the half-filled $4f^7$ orbital for $Gd^{3+}$. Therefore, insulator-to-metal transition with gigantic negative MR can occur with sufficient magnetization without the need for full moment polarization of Gd, i.e., below the spin polarization field and above the magnetic transition temperature. Additionally, such half-filled *f*-orbital yields zero orbital angular momentum for *f*-electrons and thus negligible spin-orbit coupling (SOC). Combined with the weak SOC from the light P and S atoms, it leads to extremely weak magnetic anisotropy. This is because magnetic anisotropy primarily arises from SOC and crystal field effects, while the *f*-electrons are generally well-screened from the crystal field for *Ln*-based magnetic materials. This nearly isotropic magnetism is strongly coupled with magnetotransport, giving rise to the unique isotropic colossal negative MR in GdPS[54]. Given such rare phenomenon is due to the negligible SOC in GdPS, the evolution of negative MR and insulator-to-metal transition with enhancing SOC becomes a natural question. Chemical substitution offers a viable strategy for tuning SOC. While substituting Gd with other lanthanides could enhance SOC, it would also alter the magnetic configuration, potentially weakening the strong exchange splitting essential to GdPS's unique behavior. A more targeted approach is to preserve the Gd magnetic sublattice and instead substitute the P or S sites. Since the distorted P plane is critical for breaking symmetry and opening the band gap in GdPS, chalcogen substitution, i.e., replacing S with heavier Se, emerges as a promising strategy for tuning SOC without disrupting the crystal or magnetic structure significantly.

Motivated by this, we investigated the effect of Se substitution on the magnetotransport and insulator-to-metal transition in GdPS. Our study reveals that Se substitution up to 35% indeed slightly enhances the anisotropy of the exotic magnetic transport. Additionally, Se substitution modifies the P plane, which is accompanied by a more insulating ground state and suppressed field-induced insulator-to-metal transition. This behavior likely results from an enlarged band gap in the Se-substituted samples, which cannot be easily closed by exchange splitting alone. These findings advance our understanding of the insulator-to-metal transition in GdPS and offer a pathway for tuning magnetotransport properties.

**Experimental Method**

Single crystals of GdPS$_{1-x}$Se$_x$ ($0 < x \leq 0.35$) were grown by a chemical vapor transport (CVT) method with TeCl$_4$ as the transport agent, following the synthesis procedure reported for pristine GdPS[54]. To obtain various substitutional levels, the source materials were prepared with different S-to-Se ratios. The composition and crystal structure of the resulting crystals were characterized using energy-dispersive X-ray spectroscopy (EDS) and X-ray diffraction (XRD). Powder XRD was performed on ground single crystals. Electronic transport measurements were carried out using the standard four-probe technique in a Physical Property Measurement System (PPMS, Quantum Design). Magnetization measurements were conducted with a Magnetic Property Measurement System (MPMS3, Quantum Design). High-field transport measurements were performed in a 32 T resistive magnet at the National High Magnetic Field Laboratory (NHMFL) in Tallahassee.

**Results and Discussion**

GdPS crystallizes in a layered structure consisting of flat phosphorus planes sandwiched by Gd-S layers (Fig. 1a). Although it is structurally similar to the tetragonal ZrSiS (Fig. 1a) with Si square-net planes (Fig. 1b), GdPS (Fig. 1c) undergoes an orthorhombic structure distortion where the P plane fragments into an arm-chair chain-like structure (Fig. 1d), which gaps the Dirac crossings as stated above. Composition analysis on the obtained single crystals indicates successful Se substitutions that create $GdPS_{1-x}Se_x$ where $0 < x \leq 0.35$ (Supplementary information Fig S2). The synthesis of the sample with greater Se content was challenging and the resulting crystals degrade very fast. While the pristine GdPS is relatively stable, the Se-substituted samples become more environmentally sensitive, and the samples with high amount of Se degrade when exposed to air. Samples up to $x = 0.35$ exhibit sufficient stability to support the experimental studies reported herein. Multiple measurements were performed on identical samples as well as on samples with nearly identical compositions immediately after crystal growth and phase identification to ensure reliable and reproducible properties. We did notice changes in properties when exposing these samples to air for long time, as shown in supplementary information Fig S3.

Providing the sensitivity of the electronic structure to structure distortion as noted above, we have determined the crystal structure using single crystal XRD. Interestingly, the refinement indicates a change of space group from *Pnma* for pristine GdPS to *Imma* for high-Se samples with $x \geq 0.25$, which is characterized by a modification of P-layers from arm-chair chain-like (Fig. 1d) to dimer-like (Fig. 1e). The initial attempt to refine high-Se samples isostructural to GdPS failed due to an extremely high R value and other unreasonable parameters such as atomic displacement parameters. The structure parameters extracted from the single crystal XRD refinement are listed in Table 1. It is worth noting that the structure modification by Se substitution is hardly distinguishable by powder XRD (Supplementary information Fig S1). Furthermore, though single crystal XRD can reveal the structure change, the P-P dimers may form different orders or behave differently in each P-layer, which is difficult to probe by diffraction techniques and raises the structural complexity of this system. A similar scenario has also been reported in other materials such as BaMnSb$_2$[57].

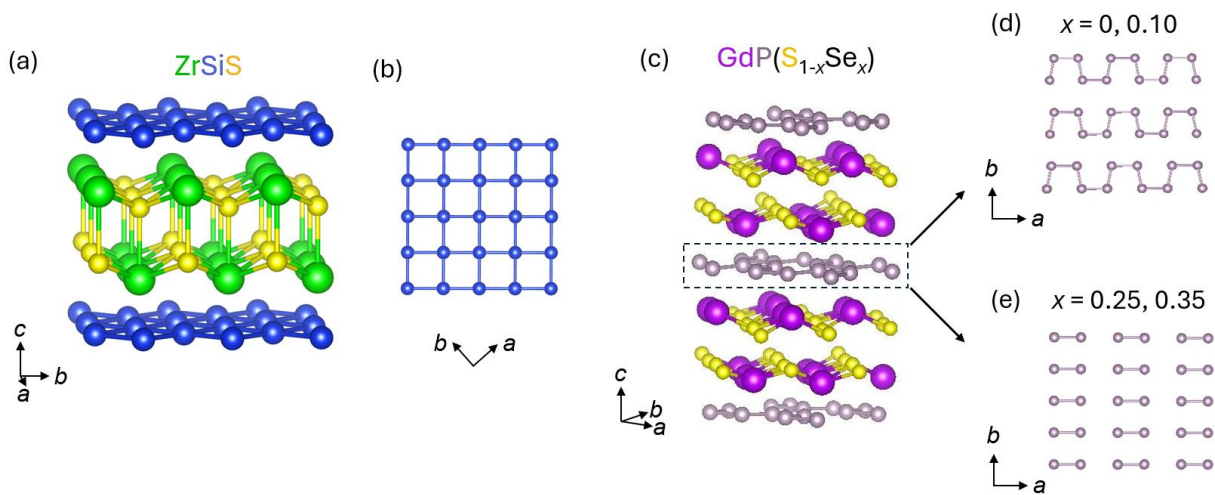

**Fig. 1.** (a) Crystal structure of ZrSiS. (b) Top view of Si square-net in ZrSiS. (c) Crystal structure of GdPS$_{1-x}$Se$_x$. The P-plane (indicated by the dashed box) for the low-Se and high-Se samples are shown as top views in (d) and (e), respectively.

**Table 1**: Lattice parameters for various GdPS$_{1-x}$Se$_x$ compositions.

|  | Space group | Lattice parameters (Å) | | | Atoms | Atomic positions | | |
| --- | --- | --- | --- | --- | --- | --- | --- | --- |
|  |  | a | b | c |  | x | y | z |
| GdPS$_{0.9}$Se$_{0.1}$ | Pnma | 5.435(0) | 5.377(7) | 16.766(5) | Gd1 | 0.4879 | 0.750 | 0.6460 |
|  |  |  |  |  | Gd2 | -0.0128 | 0.2500 | 0.6364 |
|  |  |  |  |  | P | 0.7070 | 0.5324 | 0.5018 |
|  |  |  |  |  | S1 | 0.4881 | 0.2500 | 0.6891 |
|  |  |  |  |  | S2 | -0.0096 | 0.7500 | 0.6835 |
|  |  |  |  |  | Se1 | 0.4881 | 0.2500 | 0.6891 |
|  |  |  |  |  | Se2 | -0.0096 | 0.7500 | 0.6835 |
| GdPS$_{0.75}$Se$_{0.25}$ | Imma | 5.407(6) | 5.455(9) | 16.797(6) | Gd1 | 0.5000 | 0.7500 | 0.3949 |
|  |  |  |  |  | Gd2 | 0.0000 | 0.2500 | 0.3853 |
|  |  |  |  |  | P | 0.2500 | 0.5410 | 0.2500 |
|  |  |  |  |  | S1 | 0.0000 | 0.7500 | 0.4388 |
|  |  |  |  |  | S2 | 0.5000 | 0.2500 | 0.4330 |
|  |  |  |  |  | Se1 | 0.0000 | 0.7500 | 0.4388 |
|  |  |  |  |  | Se2 | 0.5000 | 0.2500 | 0.4330 |
| GdPS$_{0.35}$Se$_{0.65}$ | Imma | 5.440(6) | 5.476(2) | 16.883(3) | Gd1 | 0.5000 | 0.7500 | 0.6064 |
|  |  |  |  |  | Gd2 | 0.0000 | 0.2500 | 0.6166 |
|  |  |  |  |  | P | 0.7500 | 0.9575 | 0.7500 |
|  |  |  |  |  | S1 | 0.0000 | 0.7500 | 0.5605 |
|  |  |  |  |  | S2 | 0.5000 | 0.2500 | 0.5666 |
|  |  |  |  |  | Se1 | 0.0000 | 0.7500 | 0.5605 |
|  |  |  |  |  | Se2 | 0.5000 | 0.2500 | 0.5666 |

As stated above, both Gd and P play key roles in the exotic properties of GdPS. The half-filled $f$-orbitals for Gd$^{3+}$ ($4f^{\,7}$) ions lead to strong magnetization as well as negligible SOC due to orbital moment quenching[54], while the arm-chair chain-like P plane breaks the symmetry which is crucial for the gapped semiconducting ground state in GdPS[54,55]. Therefore, the structure change could have a substantial impact on electronic properties. To elucidate the evolution of electronic properties with Se substitution in GdPS, we measured the temperature and field dependence of resistivity for GdPS$_{1-x}$Se$_x$ samples. As shown in Figs. 2a-d, despite a semiconducting-like behavior for all samples, GdPS$_{1-x}$Se$_x$ is more insulating with Se substitution. Though the room temperature resistivity only slightly rises from a few mΩ cm for GdPS[54] to ~10 mΩ cm for the $x = 0.35$ sample, the low temperature resistivity is significantly enhanced upon Se substitution. The ratio of resistivity at 2 K and at 300 K $\rho_{2K}/\rho_{300K}$ increases systematically from less than 10 for GdPS[54] to more than $5 \times 10^5$ for the $x = 0.35$ sample. This pronounced insulating behavior at high Se content implies an increasing bandgap, which is likely associated with the structure change as discussed above. In addition to the structure change, the Dirac nodal-line in ZrSiS-family compounds is known to be sensitive to SOC [9,11,16], while may also contribute to the bandgap opening.

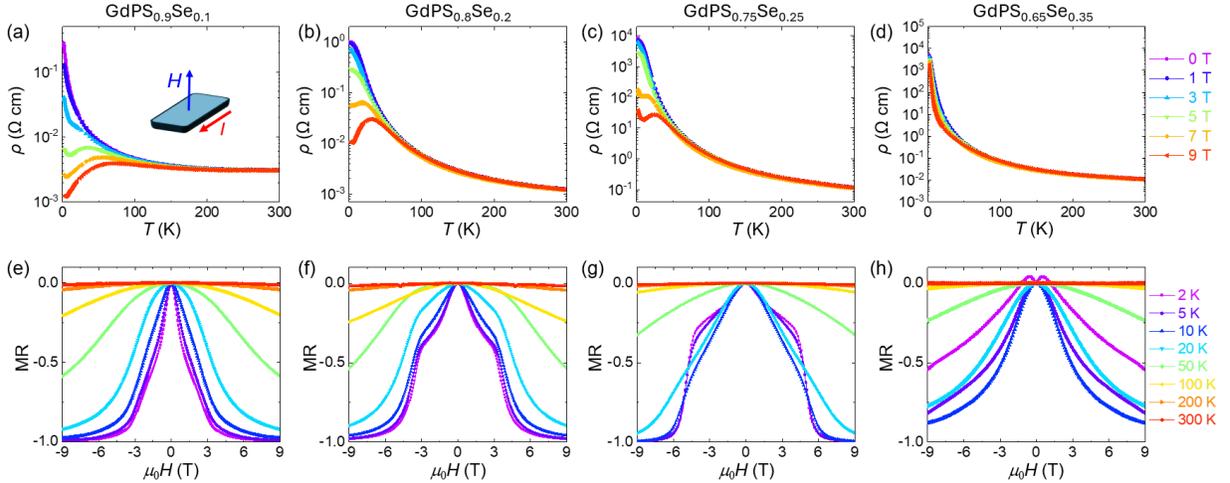

**Fig. 2.** (a-d) Temperature dependent in-plane resistivity for various GdPS$_{1-x}$Se$_x$ compositions from $x = 0.1$ to $x = 0.35$ under various magnetic fields from 0 to 9 T, applied perpendicular to the *ab*-

plane. (e-h) Normalized MR measured at various temperatures from 2 to 300 K for GdPS$_{1-x}$Se$_x$ compositions from $x = 0.1$ to $x = 0.35$.

The pristine GdPS displays an insulator-to-metal transition under magnetic field as characterized by metallic resistivity behavior, i.e., resistivity drop upon cooling[54]. Particularly, metallic temperature dependence for the entire temperature range (up to 300 K) can be achieved at a magnetic field of 9 T[54]. A similar field-induced resistivity reduction is also seen in the Se-substitute $x = 0.1$ sample, but a metallic temperature dependence only occurs at low temperatures at 9 T, following a broad resistivity hump near $T \approx 40$ K (Fig. 2a). As shown in Figs. 2b-2d, the field-induced metallicity is strongly suppressed by further Se substitution, and completely disappears in the $x = 0.35$ sample. Such suppression of field-driven metallic state is consistent with greater bandgaps in Se-substituted samples, in which the exchange splitting induced band shift is not sufficient to create a finite Fermi surface as that in pristine GdPS.

In pristine GdPS, the field-driven insulator-to-metal transition is accompanied by a gigantic negative MR that persists well above the AFM transition temperature ($T_N$) and without the need for FM polarization[54]. In Se-substituted GdPS, the negative MR remains significant. Figures 2e-h shows the MR defined as the field-induced resistivity change normalized to the zero-field resistivity value $[\rho(H)-\rho(H=0)]/\rho(H=0)$, which displays large amplitude and persists at high temperatures (50 K). Despite these common features, the $x = 0.35$ sample, which is the highest Se content available in this study, exhibits reduced MR to ~54% at 2 K and 9 T, as compared to MR value above 99% in other lower Se content samples, which should be attributed to the wider bandgap so that a field of 9 T cannot drive a metallic state. In addition, a clear dip in MR near the zero-field is seen at 2 K, which disappears with increasing temperature. Such a MR dip is reminiscent of weak anti-localization observed in materials with strong SOC[58,59], which is aligned with the substitution with heavier elements.

Large negative MR has been observed in a few materials such as EuMnSb$_2$[60] and EuTe$_2$[61,62]. The mechanism has been proposed to be correlated to the metamagnetic spin-flop transition, where net magnetization develops due to the rotation of magnetic moments from AFM to a canted configuration under a magnetic field. The gigantic negative MR in the parent compound GdPS has been revealed to arise from the emergence of a metallic phase due to band structure modification by exchange splitting that driven by magnetization[54]. Because of such a strong coupling between electronic structures and magnetism, clarifying the evolution of magnetic properties with Se substitution in GdPS can provide useful information on the evolution of magnetotransport. We have characterized the magnetic properties of GdPS$_{1-x}$Se$_x$ through temperature [$M(T)$] and field [$M(H)$] dependence of magnetization with in-plane ($H//ab$) and out-of-plane ($H//c$) magnetic fields. As shown in Fig. 3a, for both $H//ab$ and $H//c$, susceptibility for all compositions displays a sharp peak at low temperature without any irreversibility between zero-field cooling and field cooling measurements, indicating an AFM ground state that is consistent with pristine GdPS. The AFM transition temperature $T_N$ for each composition estimated from the derivative $d\chi/dT$ is summarized in Fig. 3c, which exhibits a systematic reduction with increasing Se content from 7.1 K for GdPS to $T_N \approx 6$ K for $x = 0.35$. In AFM chalcogenides, Se substitution for S can lead to both $T_N$ enhancement and reduction, depending on the nature of the dominating the magnetic exchange interactions[63]. When direct exchange interaction is dominant, $T_N$ may reduce owing to lattice expansion with Se substation. On the other hand, for compounds whose magnetic interactions occur through a super-exchange pathway, $T_N$ may increase because of the enhanced orbital overlap due to more extended orbital for Se[63]. The suppression of $T_N$ with Se substitution in GdPS appears to suggest that the direct exchange between Gd ions might be the dominant magnetic interaction in this system, which will be weakened due to the expansion of the crystal lattice upon Se substitution. However, more theoretical and experimental studies are needed to fully clarify the evolution of $T_N$.

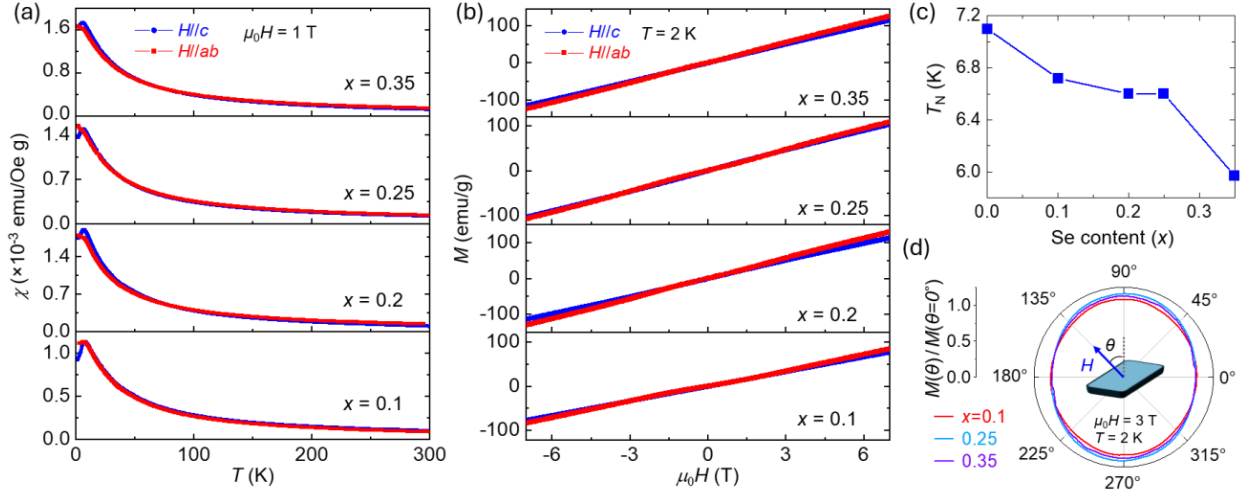

**Fig. 3.** (a) Magnetic susceptibility for GdPS$_{1-x}$Se$_x$ under out-of-plane ($H//c$) and in-plane ($H//ab$) magnetic fields of 1 T. (b) Field dependent isothermal magnetization of GdPS$_{1-x}$Se$_x$ measured under $H//c$ and $H//ab$ field orientations at 2 K. (c) Evolution of AFM transition temperature with Se content in GdPS$_{1-x}$Se$_x$. (d) Angular dependence of magnetization measured at 3 T and 2 K. The magnetization is normalized to its minimum value which occurs when the magnetic field is applied along the $c$-axis ($\theta = 0°$)

For GdPS, the weak SOC, primarily due to half-filled $f$-orbital for Gd and light elements P and S, leads to very weak magnetic anisotropy and thus the rarely seen isotropic negative MR[54]. With the heavy element Se substitution for S, the SOC is expected to enhance and lead to stronger magnetic anisotropy. To examine this effect, isothermal magnetization measurements were performed with the magnetic field applied parallel to the $ab$ plane ($H//ab$) and along the $c$- axis ($H//c$), as shown in Fig. 3b. However, these data were not sufficient for a clear quantitative assessment of anisotropy and its magnitude. We have therefore carried out angular-dependent magnetization measurements at 2 K by using a rotator to continuously change the relative angle between the sample and magnetic field, from the out-of-plane (*c*-axis) to the in-plane (*ab*-plane) direction, as illustrated in the inset of Fig. 3d. The measurements were performed at 2 K and 3 T to facilitate the direct comparison with anisotropy in magnetotransport as will be shown below. To highlight the anisotropy in magnetization, Figure 3d shows the normalized magnetization $M(\theta)/M(0°)$ in a polar plot. For the $x = 0.1$ sample, the normalized magnetization curves are nearly circular, indicating very weak anisotropy. By contrast, for the $x = 0.25$ sample, $M(\theta)/M(\theta = 0°)$ bulges toward the in-plane field direction ($\theta = 90°$ and $270°$), giving rise to an elliptically distorted polar plot that signals enhanced magnetic anisotropy. Upon further increase Se content to $x = 0.35$, the magnetic anisotropy at 3 T remains stronger than that for $x = 0.1$, but it is somewhat smaller than $x = 0.25$ sample. As shown below, such evolution of magnetic anisotropy agrees well with magnetotransport.

Figure 4 presents the field dependence of resistivity $\rho(H)$ for each composition measured at fixed temperatures from 2 to 50K and under different field orientations from out-of-plane ($H//c$, $\theta = 0°$) to in-plane ($H//ab$, $\theta = 90°$). For the lightly doped $x = 0.1$ and 0.2 samples (Figs. 4a and b), $\rho(H)$ at various field orientations almost overlap, which resembles the isotropic MR behavior in GdPS that is attributed to isotropic magnetism. In contrast, the samples with higher Se content, i.e., $x = 0.25$ and 0.35, display small but clear anisotropy at 2 K in $\rho(H)$ with varying angle $\theta$, as indicated by arrows in Figs. 4c and d. Notably, the $x = 0.25$ sample displays significantly enhanced MR anisotropy at 3 T, as indicated by the arrow in Fig. 4c. This agrees well with the enhanced magnetization anisotropy at the same field in this composition discussed above. Increasing Se content to $x = 0.35$, MR anisotropy at 3 T is less obvious (Fig. 4d), which again agrees with the reduced magnetization anisotropy in the $x = 0.35$ sample at 3 T as shown above. In this composition, strong MR anisotropy occurs around 8 T (Fig. 4d, black arrow), which is beyond the limitation (7 T) of our MPMS3 SQUID magnetometer so that the direct comparison with magnetic anisotropy is not possible. Nevertheless, the consistency of non-monotonic evolution of magnetization anisotropy and MR anisotropy at 3 T establishes a clear correlation between them. Furthermore, to better quantify the anisotropy of the MR, we define an anisotropic index as the difference between the maximum and minimum MR values obtained from angular-dependent measurements, normalized by the average MR (see Supplementary Information S4). At 2 K, samples with low Se composition exhibit an anisotropy index close to zero at all magnetic fields, indicating nearly isotropic behavior, whereas samples with higher Se content display a significantly larger anisotropy index at some magnetic fields, reflecting pronounced anisotropic behavior (Supplementary Information Fig. S4). The calculated anisotropic index reaches 0.25 at 5 T for GdPS$_{0.75}$Se$_{0.25}$ and 0.20 at 9 T for GdPS$_{0.65}$Se$_{0.35}$, whereas at other magnetic fields the anisotropic index is negligibly small.

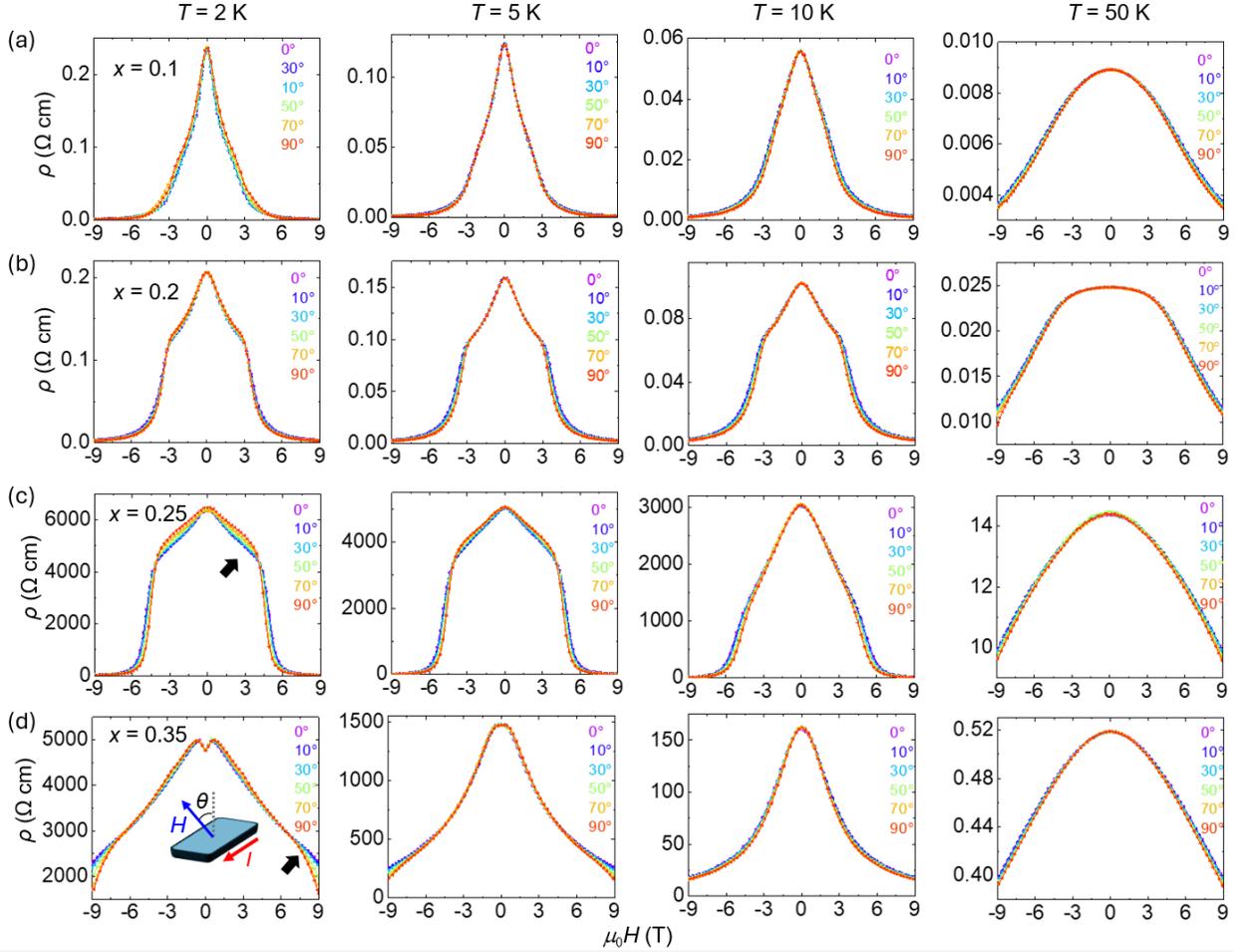

**Fig. 4.** Field dependence of resistivity of GdPS$_{1-x}$Se$_x$ (a) $x$ = 0.1 (b) $x$ = 0.2 (c) $x$ = 0.25 (d) $x$ = 0.35 at T = 2 K, 5 K, 10 K, and 50 K measured with various fixed magnetic field orientations. Inset in (d): the measurement setup.

Enhanced magnetic anisotropy is also supported by the evolution of spin polarization field. The pristine GdPS exhibit field-driven moment polarization around 15T that is almost regardless of magnetic field orientation[54]. Se-substitute compounds, though do not display quantum oscillations as the pristine GdPS, do exhibit clear signatures associated with spin polarization in high field magnetotransport. As shown in Fig. 5, resistivity drops significantly with increasing magnetic field and saturates beyond a critical field as indicated by the arrows. Such saturation of negative MR is due to the saturation magnetization in the field-polarized ferromagnetic state, where the Gd *f-d* exchange splitting reaches maxima and the electronic structure no longer changes with magnetic field[54]. At the first glance, for each composition, the spin polarization field occurs around similar fields around 15 T for both *H//c* ($\theta = 0°$) and *H//ab* ($\theta = 90°$), implying nearly isotropic spin polarization upon Se-substitution. Nevertheless, by careful comparison, one can find a systematic increase of deviation of spin polarization fields for the two field orientations with increasing Se content. Such differences between spin polarization fields, as indicated by red and bule arrows in Fig. 5, increases from ~0.7 T for the $x = 0.1$ sample to ~1.4 T for the $x = 0.2$ sample, and to ~2.6 T for the $x = 0.25$ sample. Such enhancement implies enhancing anisotropy with Se substitution. This, together with the observed lower polarization field for *H//ab*, further suggests that the system is likely to develop a more well-defined magnetic easy-axis along the ab-plane. Such development of anisotropy is most likely the consequence of the enhanced SOC with Se substitution.

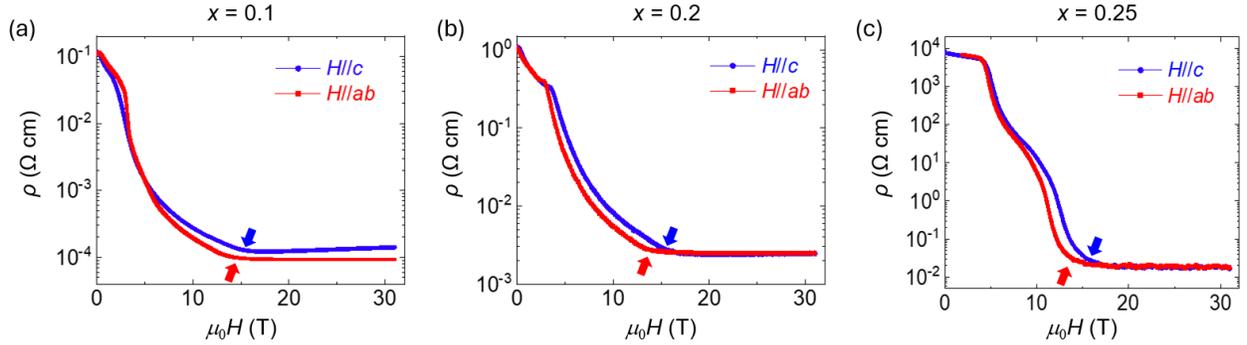

**Fig. 5.** Field dependance of resistivity of GdPS$_{1-x}$Se$_x$ at high magnetic field (a) $x = 0.1$ (b) $x = 0.2$ (c) $x = 0.25$ at $T = 1.8$ K measured for in-plane (*H//ab*, $\theta = 90°$) and out-of-plane (*H//c*, $\theta = 0°$) magnetic fields. Note that a logarithmic scale is used for resistivity for clarity.

Next, we discuss the suppression of the metal-to-insulator transition with Se substitution. As discussed above, Se substitution does not significantly impact magnetism in GdPS except for rather weak reduction of magnetic exchange interaction and some enhancement of magnetic anisotropy. The origin of the large negative MR in GdPS has been identified as an insulator-to-metal transition [54]: GdPS is AFM under zero magnetic field and the ground state is nonmetallic due to a symmetry-breaking-induced bandgap at the Fermi level. Application of a magnetic field polarizes the Gd $f$-electron moments, giving rise to strong exchange splitting that shifts the spin-polarized Gd $d$-band across the Fermi level and generates a Fermi surface. Furthermore, the isothermal magnetization is nearly isotropic with respect to the magnetic field direction, resulting in negligible anisotropy in the negative MR. These findings demonstrate that magnetism plays a central role in governing the magnetotransport properties of GdPS.

As shown in Figs. 3(a) and 3(b), the magnetization remains largely unchanged with increasing Se content, except for a modest weakening of the magnetic exchange interaction and a slight enhancement of magnetic anisotropy. Consequently, the exchange-splitting of Gd-$d$ band as a result of $f$-$d$ exchange interaction is expected to remain largely unaffected. The suppression of the insulator-to-metal transition with increasing Se content is therefore unlikely to arise from changes in magnetism. Instead, it can arise from lattice structural modifications induced by Se incorporation as revealed by XRD measurements. With the band gap enlarged due to structural modification, the band shift driven by exchange splitting is unable to reduce the gap sufficiently to enhance conductivity. As a result, the insulator-to-metal transition is suppressed. As shown above, samples with high Se content ($x$ = 0.25 and 0.35) exhibit a dimer-like structure within the P plane, which may be associated with the strongly insulating behavior observed in electronic transport measurements. Similar phenomena have been reported in structurally related materials such as NdSb$_x$Te$_{2-x}$[52] and CeSb$_{0.11}$Te$_{1.9}$[64], where modulation of the band gap is attributed to a composition-dependent tetragonal-to-orthorhombic distortion. In Se-substituted GdPS, however, fully characterizing P dimers using diffraction techniques is particularly challenging. A comparable situation occurs in BaMnSb$_2$[57] in which signature of Sb dimers were identified by scanning transmission electron microscopy and subsequently verified through the consistency between the calculated electronic band structure based on a dimer structural model and the experimentally observed bands from angle-resolved photoemission spectroscopy. Similar multimodal characterizations and theoretical calculations are needed to fully understand the atomic structure and its correlation with electronic properties in Se-substituted GdPS.

In conclusion, we have investigated the impact of Se substitution on the insulator-to-metal transition in GdPS. Our results show that Se substitution stabilizes the insulating state, suppressing the field-induced insulator-to-metal transition observed in pristine GdPS. This suppression is likely due to an increased band gap arising from structure change, which cannot be effectively closed by the exchange-splitting-induced band shift. Additionally, a slight enhancement of magnetic anisotropy induced by Se substitution is corroborated by magnetization measurements and the concomitant increase in magnetotransport anisotropy. Our findings underscore the critical role of lattice structure and magnetic anisotropy in governing the exotic transport properties of magnetic semiconductors and offer valuable insights for property engineering.

**Acknowledgement**

This work was primarily supported by the U.S. Department of Energy (DOE), Office of Science, Basic Energy Sciences program under Grant No. DE-SC0022006 (Crystal growth and low field magnetotransport). We acknowledge the MonArk NSF Quantum Foundry for magnetic property measurements using MPMS3 SQUID, which is supported by the National Science Foundation (NSF) Q-AMASE-i program under NSF Award No. DMR-1906383. High field transport was performed at the National High Magnetic Field Laboratory, which is supported by National Science Foundation Cooperative Agreement No. DMR-2128556 and the State of Florida. J. W. acknowledges NSF Award No. OSI-2328822 for structural determination using single crystal XRD and would like to thank Lee Daniels from Rigaku for the useful discussion of structure analysis. The authors are grateful to Jingyi Chen from University of Arkansas for the assistance of measurements.